\documentclass[conference]{IEEEtran}
\IEEEoverridecommandlockouts
\usepackage{cite}
\usepackage{amsmath,amssymb,amsfonts}
\usepackage{algorithmic}
\usepackage{graphicx}
\usepackage{textcomp}
\usepackage{xcolor}
\usepackage{subcaption}

\usepackage{graphicx}
\usepackage{fancyhdr}
\usepackage{listings} 
\usepackage{changepage} 
\usepackage{multirow}
\usepackage{booktabs} 

\def\BibTeX{{\rm B\kern-.05em{\sc i\kern-.025em b}\kern-.08em
    T\kern-.1667em\lower.7ex\hbox{E}\kern-.125emX}}
\begin{document}

\title{Voice Morphing: Two Identities in One Voice}

\author{\IEEEauthorblockN{Sushanta K. Pani, Anurag Chowdhury, Morgan Sandler, Arun Ross}
\IEEEauthorblockA{\textit{Michigan State University, USA} \\
rossarun@cse.msu.edu}
}

\maketitle

\begin{abstract}
In a biometric system, each biometric sample or template is typically associated with a single identity. However, recent research has demonstrated the possibility of generating ``morph" biometric samples that can successfully match more than a single identity. Morph attacks are now recognized as a potential security threat to biometric systems. However, most morph attacks have been studied on biometric modalities operating in the image domain, such as face, fingerprint, and iris. In this preliminary work, we introduce Voice Identity Morphing (VIM) - a voice-based morph attack that can synthesize speech samples that impersonate the voice characteristics of a pair of individuals. Our experiments evaluate the vulnerabilities of two popular speaker recognition systems, ECAPA-TDNN and x-vector, to VIM, with a success rate (MMPMR) of over 80\% at a false match rate of 1\% on the Librispeech dataset.
\end{abstract}

\section{Introduction}
\label{sec:intro}
Biometric systems use physical or
behavioral traits to recognize individuals  \cite{Jain2007Handbook}. A biometric system acquires a biometric sample of an individual (e.g., voice) using a sensor (e.g., microphone) and extracts a salient  feature set (or template). This template is then used to recognize the individual.  Typically, a template is associated with a single identity. However, over the past decade, several adversarial techniques, called {\em morph attacks}, have been developed to create synthetic biometric samples that can successfully match multiple identities \cite{Matteo2014Magic}.\footnote{A related vulnerability known as MasterPrint attack \cite{masterprint} or MasterFace attack \cite{masterface} has also been studied.} Furthermore, in recent times, DeepFake based synthetic image generators have been used to launch morph attacks on image-based biometric systems, viz., face, fingerprint, and iris, with high success rates \cite{venkatesh2021face}. The success of such attacks can potentially lead to compromise of security in sensitive applications where a single biometric ID card could be shared by two or more individuals for nefarious purposes. 

Existing literature on morph attacks demonstrates its potency against
biometric modalities such as face, fingerprint, and iris \cite{venkatesh2021face}, \cite{scherhag2017biometric}, \cite{sharma2021image}. For example, landmark-based \cite{Matteo2019Decoupling, Ramachandra2016Detecting} and deep learning-based \cite{zhang2021mipgan, damer2018morgan} face morph attacks have been shown to be effective against face recognition systems. Similarly, researchers have shown the possibility of launching a morph attack against iris matchers both at the image level \cite{Matteo2014Magic, sharma2021image} and feature level \cite{Ferrara2016Feasibility, Rathgeb2017Feasibility}.

The voice modality, on the other hand, has seemingly been spared from morph attacks until now. The use of voice biometrics is especially relevant in some commercial applications, such as digital voice assistants \cite{hoy2018alexa} and telephone banking \cite{melin2001ctt}. The voice morphing attack may be particularly harmful in scenarios where verification of a single identity is essential to proceed. For instance, consider an online spoken language test. In this context, the test-taking system might require the candidate to enroll their voice beforehand to ensure that the same individual appears for the test. This step is typically achieved using a speaker recognition system, designed to prevent an accomplice from taking the test on behalf of the candidate. However, with a voice morphing attack method, the candidate could enroll a morphed combination of their voice and that of an accomplice. This blend would match both identities, allowing the accomplice to take the test on the candidate's behalf by successfully matching their voice to the enrolled morphed template. This situation, coupled with the rapid adoption of voice biometric-enabled devices and services, has heightened interest in understanding their vulnerabilities to morphing attacks. Therefore, it is essential to investigate the viability and success rate of such attacks on popular speaker recognition systems.

In this paper, we propose a voice morphing technique called Voice Identity Morphing (VIM)\footnote{Note that {\em voice morphing} as defined in this work is different from previous use of this terminology in the speech literature, where it denotes modifying an individual's voice to sound like another individual.} that can synthesize artificial voice samples containing the voice characteristics of a pair of identities. Experimentally we show that the morph voice samples generated from two identities can successfully match target audio samples of both constituent identities using two different popular speaker recognition systems. The proposed method uses the DeepTalk network \cite{Chowdhury2021Deeptalk} to extract speaker embeddings from two source identities. Then, it performs a feature-level fusion of the two embeddings producing a new embedding corresponding to the morphed identity. Finally, the morphed embedding is input to a Tacotron 2-based Text-to-Speech synthesizer to generate a morphed audio sample. 

The main contributions of this preliminary work are as follows:
(a) We propose a voice identity morphing technique capable of generating speech samples that can successfully match two identities within the framework of a speaker recognition system.
(b) We evaluate and demonstrate the vulnerability of two popular speaker recognition systems, namely x-vector \cite{snyder2018x} and ECAPA-TDNN \cite{Desplanques2020Ecapa}, to our proposed method.
(c) We perform an ablation study to better understand this vulnerability, and we initiate a discussion on potential forensic measures that may counteract it.
(d) We propose directions for future study on this topic.

\begin{figure*}[htbp]
\centering
    \includegraphics[width=\linewidth]{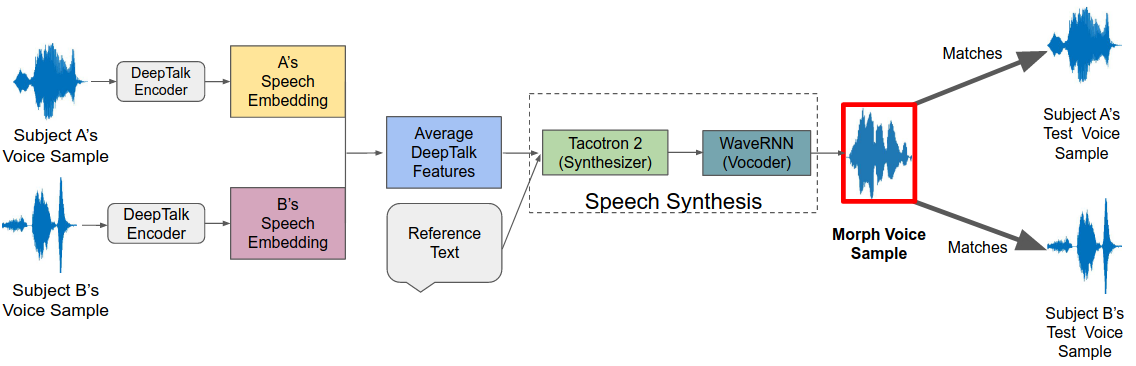}
    \caption{\label{logo} Illustration of Voice Identity Morphing: Initially, the DeepTalk encoder processes and extracts embeddings that capture the unique speaker characteristics of two distinct individuals. Subsequently, to create a morphed identity, we compute the average of these two embeddings. This averaged embedding then serves as a reference point for our speech synthesis module. Ultimately, employing this reference, the vocoder module generates a spectrogram that merges elements from both contributing speakers.}
    \label{fig:vim}
\end{figure*}

\section{Proposed Method: Voice Identity Morphing}
\label{sec:format}
Voice Identity Morphing (VIM), as shown in Figure 1, has two stages: a) synthetic voice generation and b) morph attack on a speaker recognition system. In the first stage, the proposed method generates synthetic speech samples exhibiting speaker-dependent speech characteristics pertaining to two different speakers, also referred to as the target speaker pair. The synthetic speech sample, called the morphed speech sample, is then compared to individual voice samples from the target speaker pair to launch the morph attack. An attack is successful if the morphed speech sample matches both the target speaker pair's speech samples. The morph voice generation architecture has three separate modules: Encoder, Synthesizer, and Vocoder.

  
    We use a pre-trained {\bf DeepTalk encoder} model to generate vocal style-based speaker identity embeddings of voice samples. We choose this encoder for its competitive performance with the x-vector system and its robustness to degraded audio scenarios. This encoder architecture consists of a 1D-CNN based speech filter bank also known as DeepVOX network \cite{Chowdhury2020Deepvox} and Global Style Token (GST) \cite{wang2018style} based prosody embedding network. The DeepVOX network generates short-term speaker-dependent DeepVOX features (see Table \ref{tab:deepvox} for architecture details). The GST based prosody embedding network generates a fixed dimensional reference embedding from DeepVOX features by using a 2D-CNN followed by a 128-unit GRU. The DeepTalk encoder is pre-trained on the Librispeech, VoxCeleb1 and VoxCeleb2 datasets. The synthesizer module uses these embeddings as an input during the morph sample generation stages (speech synthesis and vocoding). As an initial step of the morph sample generation stage, we average the embeddings ($Emb_{a}$ and $Emb_{b}$) of two voice samples from separate speakers to generate a morph embedding $Emb_{morph}=(Emb_{a} + Emb_{b})/2$. We perform this averaging step to incorporate features from both constituent identities. This assumes that there is an underlying geometric relationship between the identities in the learned embedding space from the DeepTalk encoder. We illustrate these relationships using t-SNE in Figure \ref{fig:tsne}.

 \begin{table}[]
     \centering
     \caption{DeepVOX network setup for learning a 40-dimension feature representation from speech frames. All rows are convolutional layers separated by a SELU activation function.}
     \begin{tabular}{c|c|c|c}
       \underline{In Channels}   & \underline{Out Channels} & \underline{Kernel} & \underline{Dilation} \\
          1 & 2 & 5x1 & 2x1 \\
          2 & 4 & 5x1 & 2x1 \\
          4 & 8 & 7x1 & 3x1 \\
          8 & 16 & 9x1 & 4x1 \\
          16 & 32 & 11x1 & 5x1 \\
          32 & 40 & 11x1 & 5x1 
     \end{tabular}
     \label{tab:deepvox}
 \end{table}
 

We use {\bf Tacotron 2 speech synthesizer} \cite{shen2018natural} to generate a mel-spectrogram for the corresponding text input. We use the Tacotron 2 synthesizer to retain consistency with the original DeepTalk architecture. Tacotron 2 architecture consists of an encoder and a decoder with an attention mechanism. The encoder creates an internal representation of input text, and the decoder uses the internal representation to generate features that encode the audio as a frame-level mel-spectrogram. The attention mechanism helps the decoder learn from the internal representation by weighting out potential failure cases where some subsequences of text are repeated or ignored by the decoder.

We use a {\bf WaveRNN-based neural vocoder} \cite{kalchbrenner2018efficient} pretrained model to generate morph samples by inverting the mel-spectrogram output from the Tacotron 2 synthesizer into audio samples. WaveRNN aims to have an expressive and non-linear transformation of the context and minimize the number of operations each step. An RNN addresses this purpose by combining the context and input within a single transformation.

\section{Experimental Protocol}
\subsection{Dataset}
We conducted experiments using the publicly accessible Librispeech dataset \cite{panayotov2015librispeech}, an audiobook corpus derived from Librivox projects. This dataset includes 1000 hours of audio data, in which, for each sample, a speaker reads English text. The dataset is divided into three subsets (100hr, 360hr, 500hr), all sampled at 16kHz. For our experiment, we utilized the 500 hour subset that consists of 1,166 participants (554 female and 612 male). We selected the 500hr subset not only because it is the largest subset, but also because it encompasses 440 speakers, each with more than 30 minutes of speaking time – a factor crucial for the morph generation process.

\subsection{Baseline Recognition Performance}
We assess speaker recognition systems' vulnerability to morph samples using two popular speaker recognition systems: x-vector \cite{snyder2018x} and ECAPA-TDNN \cite{Desplanques2020Ecapa}. We choose these systems as they are freely available and are used in a wide range of systems.\footnote{ECAPA-TDNN amassed 553,704 downloads in one month (June 2023) according to the HuggingFace website \cite{huggingface2023}} We use the implementation of these systems in Speechbrain \cite{ravanelli2021speechbrain} toolkit. The x-vector matcher is a TDNN (Time delay neural network) architecture and applies statistical pooling to extract 512-dimensional embedding for variable length utterances. The matcher utilizes categorical cross-entropy loss for training. The ECAPA-TDNN matcher architecture consists of convolutional layers, residual blocks, and attentive statistical pooling layers. It utilizes Additive Margin SoftMax Loss to generate a 192-dimensional embedding. Both matchers utilize Voxceleb1 \cite{Nagrani2017Voxceleb} and Voxceleb2 \cite{chung2018voxceleb2} datasets to train the models. They use cosine distance similarity of speaker embeddings to compare a pair of speaker identities. 

Before assessing their vulnerability, we evaluate the baseline recognition performance of these speaker recognition systems on 440 subjects in the 500-hr subset of the Librispeech dataset \cite{panayotov2015librispeech}. 
Table \ref{table:performance} provides the performance of these speaker recognition systems in terms of True Match Rate (TMR) at 1\%, 0.1\%, and 0.01\% False Match Rate (FMR). TMR is the proportion of genuine samples that were correctly matched, whereas FMR was the proportion of impostor samples that were incorrectly matched. ECAPA-TDNN model performs better than the x-vector model in correctly classifying genuine and impostor pairs.

\begin{table}[htbp]
    \caption{Performance of two speaker recognition systems in terms of TMR (\%) at 1\%, 0.1\%, and 0.01\% FMR in the Librispeech dataset. The ECAPA-TDNN and x-vector are two popular, high-performing speaker recognition systems available in the Speechbrain toolkit.}
    \label{table:performance}
    \centering
    \begin{tabular}{lccc}
        \toprule
        \multirow{2}{*}{Matcher} & \multicolumn{3}{c}{TMR (\%)} \\
        \cmidrule(lr){2-4}
        & FMR 1\% & FMR 0.1\% & FMR 0.01\% \\
        \midrule
        ECAPA-TDNN & 98.91 & 97.50 & 93.25 \\
        x-vector & 88.17 & 78.57 & 68.52 \\
        \bottomrule
    \end{tabular}
\end{table}

\subsection{Morph Generation Setup and Results}
To generate morph voice samples that incorporate both identities of two different speakers, we first fine-tune a separate Tacotron 2 synthesizer with speech samples of that speaker pair. A pre-trained Tacotron 2 synthesizer needs approximately 30 minutes of the voice samples for fine-tuning \cite{Chowdhury2021Deeptalk}. Therefore, we select 440 speakers (221 female and 219 male) which has 30 minutes or more cumulative duration of voice samples. From 440 speakers, we generate 96,580 speaker pairs ($^{440} C_{2}$ ). To generate {\it better quality} morph samples, we consider those speaker pairs which have {\it high similarity} in their speech. Each instance of Tacotron 2 takes 8-10 hours to fine-tune. Given this, we select the top 100 speaker pairs. We measure the similarity by the cosine distance of their ECAPA-TDNN-extracted speaker embeddings. Through this process, we select the top 100 speaker pairs, out of which only 43 pairs have unique speakers. The trimmed list of speaker pairs has 3 cross gender speaker pairs. Considering these 43 speaker pairs, we fine-tune 43 different Tacotron 2 synthesizers in parallel. For fine-tuning these Tacotron 2 synthesizers, we also provide 256-dimensional speaker embeddings extracted from a pre-trained DeepTalk encoder model \cite{Chowdhury2021Deeptalk} as input along with a reference text. The fine-tuned Tacotron 2 synthesizer outputs a morphed mel spectrogram which is then fed as input into the WaveRNN vocoder \cite{kalchbrenner2018efficient} to generate morphed speech samples. We create 100 such morphed samples from each speaker pair (10 samples per speaker) which results in 4,300 morphed samples. The speech samples used to generate morph samples are different from the ones used for training the Tacotron 2 synthesizer. We use the remaining voice samples of a speaker for testing. Our experiment has disjoint sets of training (60\%), morph (10\%) and test (30\%) speech samples.

To evaluate the vulnerabilities of the two speaker recognition systems against the generated morph samples (morph attack), we use the Mated Morph Presentation Match Rate (MMPMR) \cite{scherhag2017biometric} and Morphing Attack Potential (MAP) \cite{map} metrics. MMPMR is a fraction of successful morph attacks out of the total number of morph attacks. A morph attack is considered successful when the morph sample matches with test samples of both speakers. Table \ref{table:results} provides the performance of morph attacks in terms of MMPMR at different thresholds corresponding to 1\%, 0.1\%, and 0.01\% FMRs. We report the morph attack success rate in two categories: speaker pair level and morph sample level. A successful morph attack at the speaker pair level has at least one morph sample that matches the samples of both speakers. However, morph sample-level MMPMR reports the success of all morph samples irrespective of the speaker. The proposed morphing technique VIM can create morph samples attacking ECAPA-TDNN and x-vector speaker recognition systems with 95.34\% and 86.04\% respective success rates at 0.1\% FMR, for speaker pair level. The results show that the ECAPA-TDNN speaker recognition system is more susceptible to morph attacks compared to the x-vector recognition system. The considerable success rate of morph attacks could likely be related to the morph pair selection process or the effective capturing of subject information by the DeepTalk encoding method. This infers that prior knowledge of the speaker recognition system would generate stronger morph attacks. Also, we hypothesize that state-of-the-art speaker recognition systems are likely to detect vocal features of both the parent speakers in a composite audio. This may make them vulnerable to such morphing attacks as well. We find that the fusion of speech synthesis embeddings generates effective morph audio samples for use in attacks on speaker recognition systems.

\begin{table}[htbp]
    \caption{Vulnerability assessment of two speaker recognition systems to voice identity morph attack in terms of MMPMR (\%) at different threshold corresponding to 1\%, 0.1\%, and 0.01\% FMR on the Librispeech dataset.}
    \label{table:results}
    \resizebox{\columnwidth}{!}{%
    \begin{tabular}{lccc ccc}
        \toprule
        \multirow{2}{*}{Matcher} & \multicolumn{3}{c}{Speaker pair MMPMR (\%)} & \multicolumn{3}{c}{Morph sample MMPMR (\%)} \\
        \cmidrule(lr){2-4} \cmidrule(lr){5-7}
        & FMR 1\% & FMR 0.1\% & FMR 0.01\% & FMR 1\% & FMR 0.1\% & FMR 0.01\% \\
        \midrule
        ECAPA-TDNN & 100.00 & 95.34 & 81.39 & 91.23 & 62.11 & 21.58 \\
        x-vector & 93.02 & 86.04 & 9.30 & 82.13 & 38.95 & 4.32 \\
        \bottomrule
    \end{tabular}%
    }
\end{table}

\begin{table}
    \centering
    \caption{Morphing Attack Potential (MAP) \cite{map}: This metric represents the success rate (\%) of a morphed sample matching at least a specified number of probe voice samples (denoted as \# of attempts) within the Librispeech dataset, using one or both of the speaker recognition systems (SRS), namely ECAPA-TDNN and x-vector. The success rate is evaluated at three false match rate (FMR) thresholds: 1\%, 0.1\%, and 0.01\%.}
    \begin{tabular}{lcc|cc|cc}
        \toprule
        \multirow{2}{*}{\# of Attempts} & \multicolumn{2}{c}{FMR 1\%} & \multicolumn{2}{c}{FMR 0.1\%} & \multicolumn{2}{c}{FMR 0.01\%} \\
        \cmidrule(lr){2-3} \cmidrule(lr){4-5} \cmidrule(lr){6-7}
        & 1 SRS & 2 SRS & 1 SRS & 2 SRS & 1 SRS & 2 SRS \\
        \midrule
        1 & 92.0\% & 52.7\% & 60.4\% & 7.6\% & 20.2\% & 2.3\% \\
        2 & 90.2\% & 46.3\% & 54.4\% & 5.7\% & 16.5\% & 1.7\% \\
        3 & 88.9\% & 41.6\% & 50.8\% & 5.0\% & 14.3\% & 1.0\% \\
        4 & 87.9\% & 38.1\% & 47.9\% & 4.6\% & 13.0\% & 0.6\% \\
        5 & 87.0\% & 35.7\% & 45.8\% & 4.2\% & 11.4\% & 0.3\% \\
        \bottomrule
    \end{tabular}
    \label{tab:map}
\end{table}

\subsection{Result Analysis}
\begin{figure}[htbp]
  \centering
  \begin{subfigure}[b]{0.49\columnwidth}
    \includegraphics[width=\linewidth]{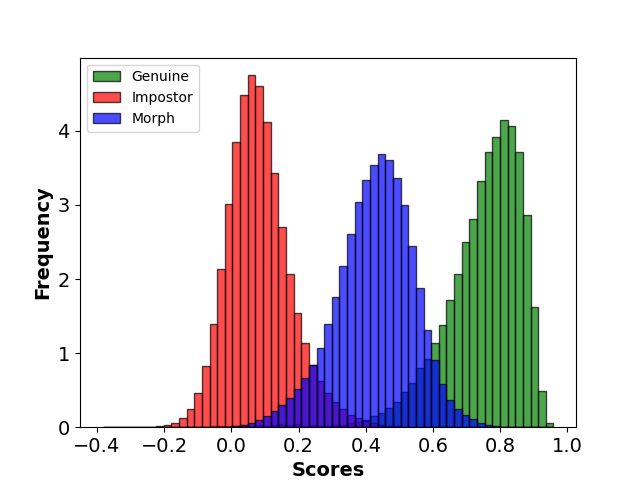}
    \caption{ECAPA-TDNN}
    \label{subfig:ecapa}
  \end{subfigure}
  \begin{subfigure}[b]{0.49\columnwidth}
    \includegraphics[width=\linewidth]{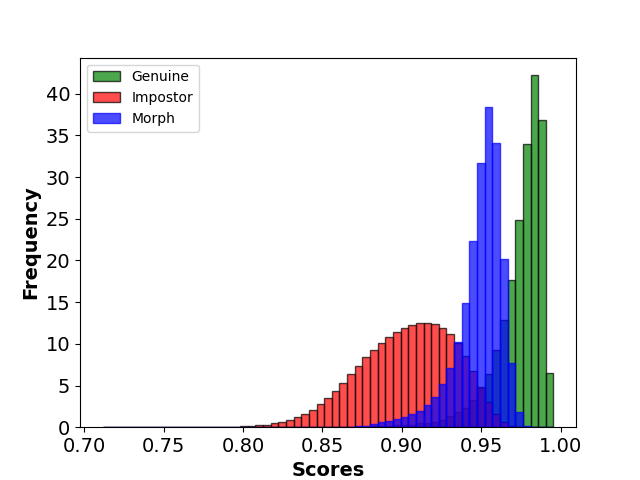}
    \caption{x-vector}
    \label{subfig:xvector}
  \end{subfigure}
  \caption{Speaker recognition match score distributions of non-morph versus non-morph genuine (Green), non-morph versus non-morph impostor (Red), and morph versus non-morph genuine morph scores (Blue) using ECAPA-TDNN and x-vector embeddings.}
  \label{fig:gim}
\end{figure}

We further analyze our morph attack performance using: 1) histogram plots, 2) t-SNE plots, and 3) morphing attack potential (MAP). Figure \ref{fig:gim} shows the histogram plots of match scores corresponding to genuine pairs (green), impostor pairs (red), and pairs which include at least one morphed sample (blue) for both speaker recognition systems. In both systems, we find that the morphed pairs match score distribution lies between genuine and impostor score distributions. Morph samples are classified as genuine matches in the ECAPA-TDNN and x-vector systems with recognition thresholds of 0.46 and 0.96 respectively at 0.1\% FMR.   

\begin{figure}[]
         \centering
          \includegraphics[width=7cm]{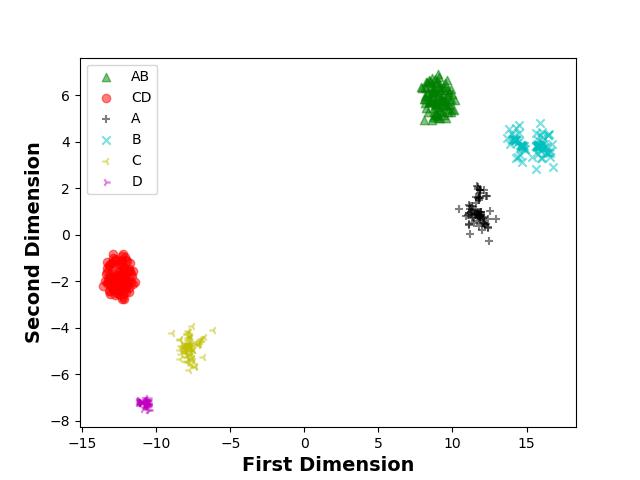}
     \caption{t-SNE plot which illustrates high-dimensional ECAPA-TDNN embeddings of morph speech samples from two separate speaker pairs (AB and CD) and non-morph speech samples of individual speakers (A, B, C, and D). Morph embeddings of each pair are closer to the non-morph embeddings of their constituent speakers.}
     
    \label{fig:tsne}
    \end{figure}

The second analysis we perform is based on the t-SNE dimensionality reduction technique. The t-SNE \cite{van2008visualizing} method helps visualize high-dimensional embeddings in a two-dimensional space by reducing the dimension. Figure \ref{fig:tsne} shows the t-SNE plot of morph sample embeddings from two speaker pairs (AB and CD) along with non-morph samples of four constituent speakers (A, B, C, and D). The embeddings are extracted by the ECAPA-TDNN recognition system. Here, embeddings of morph samples of one speaker pair (AB) are closer to embeddings of A and B speakers. Similarly, embeddings of morph samples of another speaker pair (CD) are closer to embeddings of C and D speakers. The analysis again validates the effectiveness of the proposed morphing technique and the potential threat of morph attacks.

The Morphing Attack Potential (MAP) \cite{map} constitutes the third analysis. This metric takes into account multiple Speaker Recognition Systems (SRS) to ensure generality, and a variable number of verified probe samples for robustness. The result is a matrix (Table \ref{tab:map}) in which one axis represents the number of probe samples (referred to as the number of attempts), and the other axis represents the number of SRS. The entries in each row represent the success rate (in percentage) of a morphed sample matching at least a specified number of probe voice samples (referred to as the number of attempts) using either or both of the SRS, viz., ECAPA-TDNN and x-vector. We report the success rates over three FMR thresholds of 1\%, 0.1\%, and 0.01\%. The results imply that VIM is effective at a fairly competitive FMR of 1\%, but suggest there is still room for improvement in performance at very low FMR thresholds. This may be attributed to the morph selection process or perhaps to the pre-trained models used in the encoder and speech synthesis steps.

\section{Summary and Future Work}
To the best of our knowledge, this preliminary work is the first to demonstrate the vulnerability of speaker recognition systems to morph attacks. In this regard, we propose a voice morphing technique called VIM to generate speech samples corresponding to the identities of two subjects. Using these morph samples, we demonstrate a morph attack success rate of over 80\% on two popular speaker recognition systems (ECAPA-TDNN and x-vector). 
As future work, we propose to select high-similarity pairs for a morphing attack using x-vector to investigate whether the selection process plays a vital role in the performance of such an attack. Additionally, evaluating newer speaker recognition systems such as TitaNet \cite{titanet} and MFA-Conformer \cite{mfa} would provide more insight into the generalizability of VIM. Comparing other speech synthesis systems in the speech synthesis step would shed light on the role this step plays in the VIM attack. Furthermore, we aim to develop a system for detecting morphed speech samples, possibly through the identification of their constituent identities. It may also be interesting to explore the maximum number of identities that can be combined into a single audio sample using VIM.

\section{Reproducibility}
The code for generating VIM samples can be found online at our Github link. \footnote{https://github.com/morganlee123/VIM}

\bibliographystyle{IEEEtran}
\bibliography{main}

\end{document}